\documentstyle[aps,floats,psfig,epsf,twocolumn]{revtex}
\draft

\begin{document}
\title{Order parameter symmetries for magnetic and superconducting
instabilities: Bethe-Salpeter analysis of functional renormalization-group
solutions}
\author{A. A. Katanin$^{a,b}$ and A. P. Kampf$^c$}
\address{$^a$Max-Planck-Institut f\"ur Festk\"orperforschung, D-70569
Stuttgart, Germany\\
$^b$Institute of Metal Physics, 620219 Ekaterinburg, Russia\\
$^c$Theoretische Physik III, Elektronische Korrelationen und Magnetismus,\\
Institut f\"ur Physik, Universit\"at Augsburg, D-86135 Augsburg, Germany}
\address{~\\
\parbox{14cm}{\rm \medskip \vskip0.2cm The Bethe-Salpeter equation
is combined with the temperature-cutoff functional renormalization
group approach to analyze the order parameter structure for the
leading instabilities of the 2D $t$-$t^{\prime}$ Hubbard model. We find
significant deviations from the conventional s-, p-, or d-wave forms, which is
due to the frustration of antiferromagnetism at small and intermediate
$t^{\prime}$. With adding a direct antiferromagnetic spin-exchange
coupling the eigenfunctions in the particle-hole channel have extended
s-wave form, while in the particle-particle singlet pairing channel a higher
angular momentum component arises besides the standard d$_{x^2-y^2}$-wave
component, which flattens the angular dependence of the gap. For $t^{\prime}$
closer to $t/2$ we find a delicate competition of ferromagnetism and triplet
pairing with a nontrivial pair-wavefunction.
 }}
\maketitle

\tighten

It is by now well established that the superconducting order parameter in
high-T$_c$ compounds is well described by a ($\cos k_x-\cos k_y$)-momentum
dependence: it is largest at the Fermi surface (FS) points close to $(\pi,0)$
and $(0,\pi)$ and vanishes at the FS crossings on the Brillouin zone (BZ)
diagonals. Accurate measurements of the gap function, however, revealed a
slight deviation from this d$_{x^2-y^2}$-wave momentum dependence \cite{Mesot},
with a flatter angular dependence near the nodal
points. The symmetry and the details of the momentum dependence of the
superconducting order parameter are closely related to the structure of the
effective attractive interaction between the electrons. The precise momentum
dependence of the energy gap therefore contains valuable information about
the underlying pairing mechanism, for which antiferromagnetic (AFM) spin
fluctuations are a viable candidate for cuprates \cite
{Scalapino,BSW,Pines,Chubukov}.

Another type of unconventional superconductor is the layered ruthenate
Sr$_2$RuO$_4$ \cite{PTReview}, which most likely has triplet pairing with
p-wave symmetry \cite{SrRuGap}. It was proposed that the pairing in this
material results from ferromagnetic (FM) spin fluctuations
\cite{Mazin,Murakami}. Although inelastic neutron scattering has so far been
unsuccessful to detect significant low-energy FM spin fluctuations
\cite{Maeno}, this idea finds support from the enhanced tendencies towards
ferromagnetism in the electron doped compound Sr$_{2-x}$La$_x$RuO$_4$
\cite{ED} and in isoelectronic Ca$_2$RuO$_4$ under hydrostatic pressure
\cite{Maeno1}.

The important role of AFM and FM fluctuations as a possible driving source
of singlet and triplet superconductivity, respectively, was emphasized
early on in the pioneering work in Refs. \cite{KL,BS,SLH}. Recently, the
interplay of antiferromagnetism and d-wave superconductivity (dSC) and
ferromagnetism and p-wave superconductivity (pSC), respectively, was
reconsidered within the $t$-$t^{\prime }$ Hubbard model using functional
renormalization-group (fRG) techniques
\cite{Zanchi,Metzner,SalmHon,SalmHon1,KK}. Early versions of fRG
\cite{Zanchi,Metzner,SalmHon}, which used the momentum cutoff procedure, were
unable to search for ferromagnetism. This drawback is overcome in the
temperature-cutoff fRG approach (TCRG) \cite{SalmHon1}, which proved
successful in describing both, AFM and FM instabilities together with
singlet- and triplet superconducting pairing in the weak-coupling
$t$-$t^{\prime}$ Hubbard model and its extensions in a broad parameter range
\cite{SalmHon1,KK,KK1}. In the previous fRG analyses it has been a common
practice to assume order parameter structures, which have the form of square
lattice basis functions with s-, p- or d-wave symmetry
\cite{Zanchi,Metzner,SalmHon,SalmHon1,KK,KK1}. Although it was recognized that
in the RG flow of the order parameter susceptibilities also the momentum
dependence acquires specific corrections to their initial form \cite
{Metzner,SalmHon1,KK,KK1,Neumayr}, these corrections were so far not
analyzed in detail. Indeed, actual order parameter structures may have
admixtures of different symmetry components, and it is necessary to develop
an unbiased method to determine their precise momentum dependence.

In the present paper we use the Bethe-Salpeter equations to extract
eigenfunctions and eigenvalues of the effective interaction in the
particle-particle (pp) or the particle-hole (ph) channel -- similarly as in
previous quantum Monte Carlo studies \cite{BetheSal}. Here we consider a
combination of the Bethe-Salpeter equations and the fRG approach. We choose
the TCRG version \cite{SalmHon1} as the most suitable tool, because the
effective model obtained within this scheme does not contain any
unintegrated degrees of freedom even at the intermediate stages of the RG
flow.

We apply this procedure to identify the order parameter structure for the
leading instabilities of the 2D $t$-$t^{\prime }$ extended Hubbard model
$H_\mu=H-(\mu-4t^{\prime})N$ with
\begin{equation}
H=-\sum_{ij\sigma }t_{ij}c_{i\sigma }^{\dagger }c_{j\sigma
}+U\sum_in_{i\uparrow }n_{i\downarrow }+J\sum_{\langle ij\rangle }{\bf S}
_i\cdot {\bf S}_j\,,  \label{H}
\end{equation}
where $t_{ij}=t$ for nearest neighbor (nn) sites $i$ and $j$ and
$t_{ij}=-t^{\prime }$ for next-nn sites ($t,t^{\prime }>0$) on a square
lattice; for convenience we have shifted the chemical potential $\mu$ by
$4t^{\prime}$. In Eq. (\ref{H}) we have included a direct nn spin exchange
interaction $J$; ${\bf S}_i=c_{i\alpha}^{\dagger}\mbox{\boldmath$\sigma$}_{
\alpha \beta }c_{i\beta }/2,$ and $\mbox{\boldmath$\sigma$}$ denotes the
Pauli matrices. While such an interaction is generated from the on-site
Coulomb repulsion at strong-coupling, we add it here as an independent
interaction in the weak-coupling regime, where the RG scheme is applicable.

We follow the many-patch fRG version for one-particle irreducible Green
functions as proposed in Ref. \cite{SalmHon1}. This TCRG scheme uses the
temperature as a natural cutoff parameter and accounts for excitations with
momenta far from and close to the FS, which is necessary for the description
of instabilities arising from zero-momentum ph scattering, e.g.
ferromagnetism. Neglecting the frequency dependence of the vertices, which is
expected to have minor relevance in the weak-coupling regime, the RG
differential equation for the interaction vertex has the form
\cite{SalmHon1}
\begin{eqnarray}
\frac{{\rm d}V_T}{{\rm d}T}=-V_T\circ \frac{{\rm d}L_{{\rm pp}}}{{\rm d}T}%
\circ V_T+V_T\circ \frac{{\rm d}L_{{\rm ph}}}{{\rm d}T}\circ V_T\,,
\label{dV}
\end{eqnarray}
where $\circ $ is a short notation for summations over intermediate momenta
and spin,
\begin{equation}
L_{\text{ph,pp}}({\bf k},{\bf k}^{\prime })=\frac{f_T(\varepsilon _{{\bf k}})
-f_T(\pm \varepsilon _{{\bf k}^{\prime }})}{\varepsilon _{{\bf k}}\mp
\varepsilon _{{\bf k}^{\prime }}},
\label{Lphpp}
\end{equation}
and $f_T(\varepsilon )$ is the Fermi function. The upper sign in Eq.
(\ref{Lphpp}) is for $L_{ph}$ and the lower sign for $L_{pp}$, respectively.
Eq. (\ref{dV}) has to be solved with the initial condition
$V_{T_0}({\bf k}_1,{\bf k}_2,{\bf k}_3,{\bf k}_4)=U$; the initial temperature
is chosen as large as $T_0=400t$.

We discretize the momentum space in $N_p=48$ patches using the same patching
scheme as in Ref. \cite{SalmHon1}. This reduces the integro-differential
equations (\ref{dV}) and (\ref{dH}) to a set of 5824 differential equations,
which were solved numerically. The evolution of the vertices with decreasing
temperature determines the temperature dependence of the susceptibilities
according to \cite{SalmHon1,SalmHon02}
\begin{eqnarray}
\displaystyle{\frac{{\rm d}\chi _{m,r}}{{\rm d}T}} &=&\sum_{{\bf k}^{\prime
}}{\cal R}_{{\bf k}^{\prime }}^{m,r}{\cal R}_{\mp {\bf k}^{\prime }+{\bf q}%
_m}^{m,r}\displaystyle{\frac{{\rm d}L_{\text{pp},\text{ph}}({\bf k}^{\prime
},\mp {\bf k}^{\prime }+{\bf q}_m)}{{\rm d}T}},  \label{dH} \\
\displaystyle{\frac{{\rm d}{\cal R}_{{\bf k}}^{m,r}}{{\rm d}T}} &=&\mp \sum_{%
{\bf k}^{\prime }}{\cal R}_{{\bf k}^{\prime }}^{m,r}\Gamma _m^T({\bf k},{\bf %
k}^{\prime })\displaystyle{\frac{{\rm d}L_{\text{pp},\text{ph}}({\bf k}%
^{\prime },\mp {\bf k}^{\prime }+{\bf q}_m)}{{\rm d}T}},  \nonumber
\end{eqnarray}
where ($\Gamma _m^T\equiv \Gamma _m^T({\bf k},{\bf k}^{\prime })$)
\begin{equation}
\Gamma _m^T=\left\{
\begin{array}{cl}
\begin{array}{c}
V_T({\bf k},{\bf k}^{\prime },{\bf k}^{\prime }+{\bf q}_m) \\
-2V_T({\bf k},{\bf k}^{\prime },{\bf k}+{\bf q}_m)
\end{array}
& \text{phc (PI, DDW)}, \\
V_T({\bf k},{\bf k}^{\prime },{\bf k}^{\prime }+{\bf q}_m) & \text{phs (AFM,
FM),} \\
V_T({\bf k},-{\bf k+q}_m,{\bf k}^{\prime }) & \text{pp (dSC, pSC).}
\end{array}
\right.   \label{Gamma}
\end{equation}
$m$ denotes one of the possible channels: ph spin (phs), which traces FM and
AFM instabilities, ph charge (phc) for the Pomeranchuk instability (PI) \cite
{Pomeranchuk} or d-density wave (DDW) \cite{DDW}, or pp for superconducting
singlet and triplet instabilities; ${\bf q}_m={\bf Q=}(\pi ,\pi )$ for AFM
and DDW and ${\bf q}_m={\bf 0}$ otherwise. In the following we use the
notation $m{\bf q}_m$ to denote the specific channel. The index $r$ in Eq.
(\ref{dH}) represents the symmetry of the corresponding channel. The upper
signs and pp-indices in Eq. (\ref{dH}) refer to the superconducting
instabilities (dSC and pSC), and the lower signs and ph-indices refer to the
charge and magnetic instabilities.

The three-point vertices ${\cal R}_{{\bf k}}^{m,r}$ can be considered
as vertices describing the propagation of the electron in a static external
field; their diagrammatic representation
is shown in Fig. 1. The initial conditions at $T_0$ for Eqs. (\ref{dH}) are
${\cal R}_{{\bf k}}^{m,r}=f_{{\bf k}}^{(r)}$ and $\chi_{m,r}=0$, where the
function $f_{{\bf k}}^{(r)}$ belongs to one of the irreducible representations
of the point group of the square lattice ($A_{1g}$ or $B_{1,2g}$), e.g.

\begin{figure}[t!]
\psfig{file=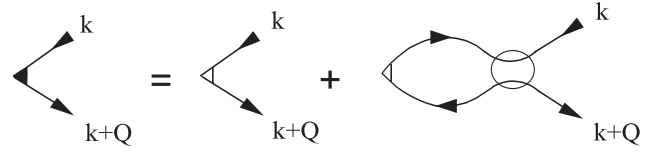,width=90mm,silent=} \vspace{2mm}
\caption{Diagrammatic representation of the three-point vertex
${\cal R}_{{\bf k}}^{m,r}$ (solid triangle), which has the structure of a
vertex diagram in a static external field acting in the channel m with the
r-th initial vertex function $f_{{\bf k}}^{(r)}$ defined in Eq. (\ref{ff}) and
shown as an open triangle. The specific example shows the phs{\bf Q} channel.
The circle represents the reducible electron-electron interaction vertex.}
\label{fig:Fig1}
\end{figure}

\begin{equation}
f_{{\bf k}}^{(r)}=A^{-1}\left\{
\begin{array}{cll}
\cos k_x-\cos k_y & \text{d$_{x^2-y^2}$-wave } & (B_{1g}), \\
\sin k_{x(y)} & \text{p-wave} & (B_{2g}), \\
1 & \text{s-wave} & (A_{1g}) \label{ff}
\end{array}
\right.
\end{equation}
with a normalization coefficient $A=(1/N)\sum_{{\bf k}}f_{{\bf k}}^2$. The
momentum dependence of the vertices ${\cal R}_{{\bf k}}^{m,r}$ changes during
the RG flow and at low temperatures it is expected to reflect the momentum
dependence of the ground-state order parameter of the selected symmetry.
However, these vertex functions depend on the initial choice of the functions
$f_{{\bf k}}^{(r)}$, and therefore they can not serve as an unbiased tool to
obtain the structure of the ground-state order parameters.

To perform an alternative analysis, which is not based on the knowledge of
the wave functions at high temperatures, we consider the solution of the
Bethe-Salpeter equations \cite{BetheSal}
\begin{eqnarray}
\sum_{{\bf p}}\overline{\Gamma }_{ph}({\bf k},{\bf p};{\bf p},{\bf k})L_{ph}(
{\bf p},{\bf p}+{\bf q}_m)\phi _{{\bf p}}^{ph} &=&\lambda _{ph}\phi _{{\bf k}
}^{ph}\,,  \nonumber \\
-\sum_{{\bf p}}\overline{\Gamma }_{pp}({\bf k},{\bf p};{\bf p},{\bf k}
)L_{pp}({\bf p},{\bf p})\phi _{{\bf p}}^{pp} &=&\lambda _{pp}\phi _{{\bf k}
}^{pp}\, ,
\label{BSirr}
\end{eqnarray}
where $\overline{\Gamma }_{ph}$ and $\overline{\Gamma }_{pp}$ denote
irreducible vertices in ph and pp channels, respectively. Exploiting the
connection to reducible vertices, Eqs. (\ref{BSirr}) can be rewritten as
\cite{BetheSal}
\begin{equation}
\sum_{{\bf p}}\Gamma _m^T({\bf k},{\bf p})L_{ph,pp}^T({\bf p},\pm {\bf p}+
{\bf q}_m)\phi _{{\bf p}}^{m,l}=\displaystyle{\frac{\lambda _{m,l}\phi_{{\bf
k}}^{m,l}}{1-\lambda _{m,l}}}\, ,
\label{BSr}
\end{equation}
where $l$ enumerates eigenvalues and -functions for a given channel $m$. The
reducible vertices $\Gamma _m^T({\bf k},{\bf p})$ can be directly extracted
from the fRG flow according to Eq. (\ref{Gamma}).

The value $\lambda _{m,l}=1$ corresponds to an ordering instability with the
symmetry of the eigenfunction $\phi _{{\bf k}}^{m,l}$. Therefore, tracing
the temperature dependence of eigenvalues and -functions allows to identify
both, the leading instabilities {\it and} their concomitant order parameter
structure. We stop the fRG flow at the temperature $T_X$, where the maximum
interaction vertex $V_{\max}\equiv\max\{V({\bf k}_1,{\bf k}_2;{\bf k}_3,
{\bf k}_4)\}$ reaches the value $20t$.
$T_X$ should be understood as a crossover temperature into a renormalized
classical regime with exponentially large correlation length \cite{KK1},
we have verified that the following results for the eigenfunctions are
only weakly dependent on the choice of $T_X$.

Eigenfunctions and eigenvalues of the Bethe-Salpeter equations provide more
detailed information than can be obtained from the momentum dependence of the
three-point vertices ${\cal R}_{{\bf k}}^{m,r}$, which determine the
order-parameter susceptibilities according to Eq. (\ref{dH}). Unlike the
vertices ${\cal R}_{{\bf k}}^{m,r}$, the solutions of the Bethe-Salpeter
equations do not depend on the initial choice of the functions $f_{{\bf k}}^{
(r)}$ and, therefore, do not require {\it a priori} the knowledge of the
symmetry of the leading instability.

\begin{figure}[t!]
\psfig{file=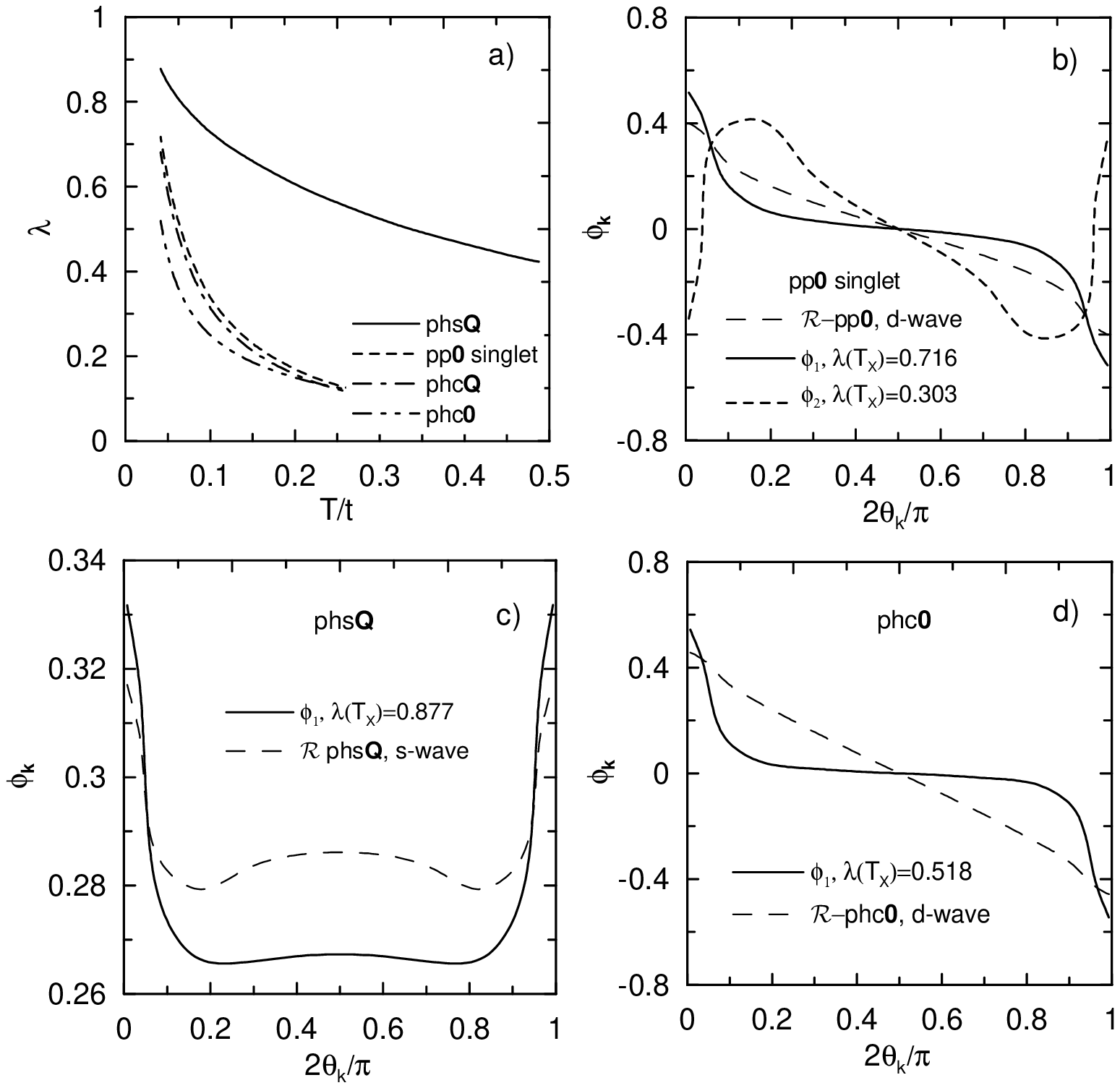,width=90mm,silent=} \vspace{2mm}
\caption{Eigenvalues (a) and angular $\theta_{{\bf k}_F}$-dependence on the
FS of the eigenfunctions $\phi _{{\bf p}}^{m,l}=\phi _l$ of the
Bethe-Salpeter equation and the normalized three-point vertices ${\cal R}_{
{\bf k}}^{m,r}$ (denoted as ${\cal R}$-$m,r$) at $T=T_X$ (b,c,d) for
$t^{\prime }=0.1t,$ $U=2t,$ $J=\mu=0$. $T_X$ is the lowest temperature
reached in the fRG flow (see (a)).}
\label{fig:Fig2}
\end{figure}

The three-point vertices
\begin{equation}
\widetilde{{\cal R}}_{{\bf k}}^{m,r}=f_{{\bf k}}^{(r)}+\sum_{{\bf p}}\Gamma
_m^T({\bf k},{\bf p})L_{ph,pp}^T({\bf p},\pm {\bf p}+{\bf q}_m)f_{{\bf p}}^{
(r)}
\end{equation}
described by the diagram of Fig. 1 can be obtained directly from the solutions
of the Bethe-Salpeter equations. Expanding the functions $f_{{\bf k}}^{(r)}$
in terms of eigenfunctions of the Bethe-Salpeter equation and using Eqs.
(\ref{BSr}), we obtain
\begin{equation}
\widetilde{{\cal R}}_{{\bf k}}^{m,r}=\sum_l\frac{\phi _{{\bf k}}^{m,l}}{%
1-\lambda _{m,l}}\sum_{{\bf k}^{\prime }}f_{{\bf k}^{\prime }}^{(r)}\phi _{%
{\bf k}^{\prime }}^{m,l}\, .
\label{RR}
\end{equation}
The vertices $\widetilde{{\cal R}}_{{\bf k}}^{m,r}$ found in this way,
however, do not necessarily coincide with those obtained directly from the
RG procedure, Eq. (\ref{dH}), because generally the one-loop approximation
does not exactly reproduce the solution of the corresponding Bethe-Salpeter
equation. The only case when the three-point vertices ${\cal R}_{{\bf k}}^{m,
r}$ and $\widetilde{{\cal R}}_{{\bf k}}^{m,r}$ coincide is the ladder
approximation when either only the pp-channel or one of the ph-channels in
Eq. (\ref{dV}) is retained. Nevertheless, as we will see below, the results
for the vertex ${\cal R}_{{\bf k}}^{m,r}$ from Eqs. (\ref{dH}) are
qualitatively similar to those found from Eq. (\ref{RR}).

It is clearly established from Eq. (\ref{RR}) that $\widetilde{{\cal R}}_{{\bf
k}}^{m,r}$ (and similarly ${\cal R}_{{\bf k}}^{m,r}$) is actually a mixture of
different eigenfunctions $\phi_{{\bf k}}^{m,l}$, whose weights are
proportional to $1/(1-\lambda _{m,l})$ with coefficients determined by the
overlap of the eigenfunction with $f_{{\bf k}^{\prime}}^{(r)}$. If one of the
eigenvalues $\lambda _{m,l}$ is much closer to unity than all the others, the
corresponding term in the sum over $l$ in Eq. (\ref{RR}) is expected to
dominate and $\widetilde{{\cal R}}_{{\bf k}}^{m,r}$ essentially coincides with
the corresponding eigenfunction $\phi_{{\bf k}}^{m,l}$.

Below we discuss the results of the numerical solution of the Bethe-Salpeter
equations and compare them to the results for the three-point vertices
${\cal R}_{{\bf k}}^{m,r}$ for different parameter regimes. We start in Fig.
2 with results at the van Hove (vH) band filling ($\mu =0$) for small
$t^{\prime }=0.1t$, $J=0$, and $U=2t$. The AFM phs{\bf Q} instability has the
largest eigenvalue at $T_X$ in agreement with previous fRG work based on the
analysis of order parameter susceptibilities \cite{SalmHon1}. The
corresponding eigenfunction (Fig. 2c) has a slight variation around the FS
with an enhancement near ($\pi ,0$) and ($0,\pi$), which most likely
originates from the vH singularity nature of these points. The
eigenfunctions in the subleading phc{\bf 0} and pp{\bf 0} channels are
sizable near ($\pi ,0$) and ($0,\pi ,$) only. Although the eigenvalue of the
zero-momentum ph instability in the charge channel (phc{\bf 0}) is
relatively small at $T=T_X$ ($\lambda \simeq 0.5$), it rapidly increases at
low temperatures. The corresponding eigenfunction (Fig. 2d) is antisymmetric
with respect to 90$^{\circ }$ rotation, thus providing the possibility for
the Pomeranchuk instability with a spontaneous d-wave like deformation of
the FS \cite{Yamase,Metzner1}. The eigenvalues for the PI at $t^{\prime
}=0.1t$ are larger than those at $t^{\prime}=0$ (not shown), which supports
the conclusion of Ref. \cite{Metzner1}, that a finite value of $t^{\prime}$
at vH band fillings enhances the tendency towards a Pomeranchuk instability.

The momentum dependence of the three-point vertices ${\cal R}_{{\bf k}}^{m,r}$
is qualitatively similar to the leading eigenfunctions in the pp{\bf 0} and
phs{\bf Q }channels, however with a smaller variation around the FS. In the
phc{\bf 0} channel we observe a stronger difference of the vertex
${\cal R}_{{\bf k}}^{phc,d\text{-wave}}$, which has almost the
d$_{x^2-y^2}$-wave form, from the corresponding eigenfunction of the
Bethe-Salpeter equation.

\begin{figure}[t!]
\psfig{file=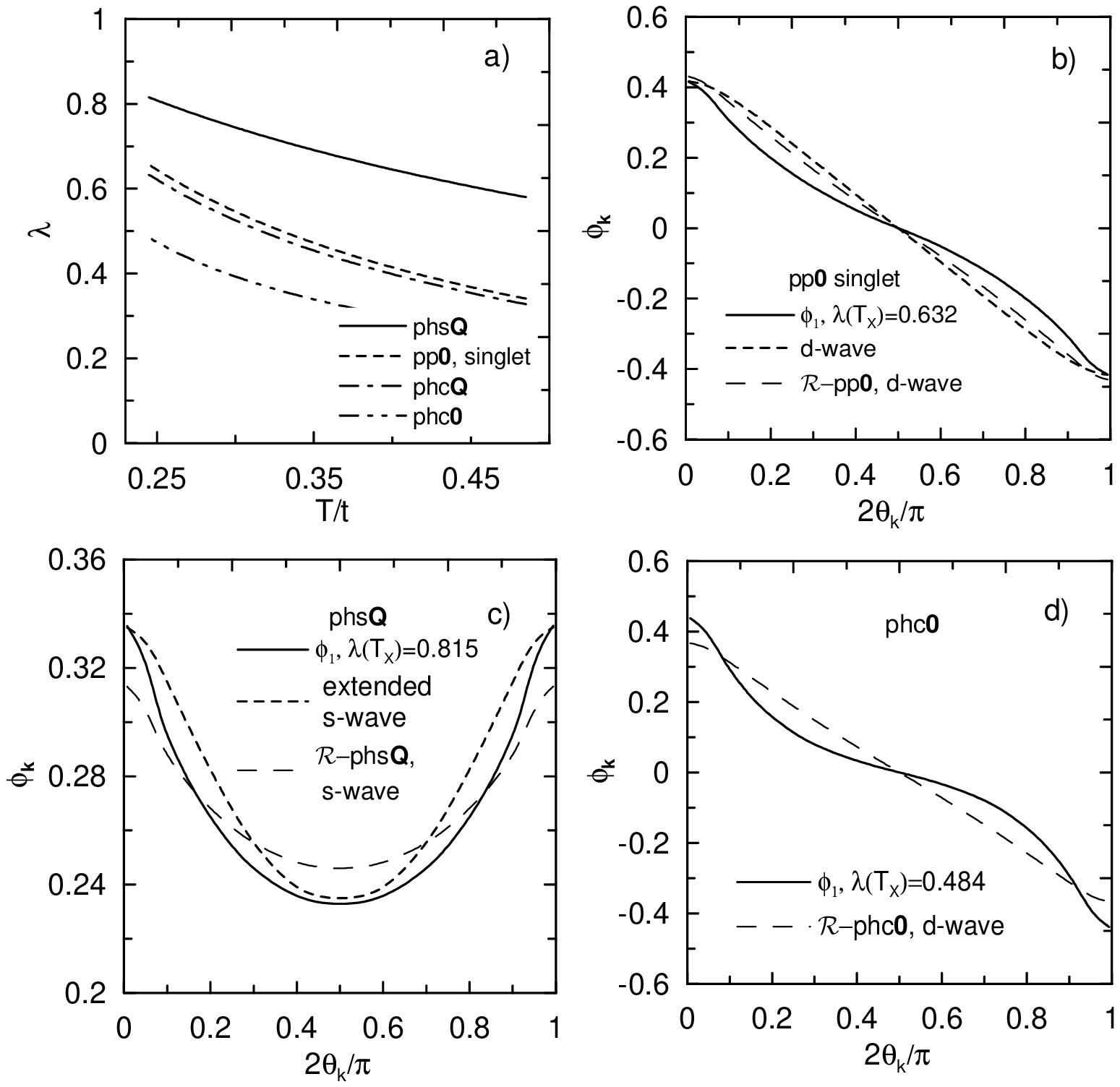,width=90mm,silent=} \vspace{2mm}
\caption{Same as in Fig. 1 for $U=1.5t,$ $J=0.3t.$}
\label{fig:Fig3}
\end{figure}

To investigate the role of $t^{\prime }$-induced frustration of
antiferromagnetism, we turn on a small direct exchange interaction $J=0.3t$
to strengthen the AFM correlations, and decrease simultaneously the value of
$U$ to $1.5t$ in order to remain in the weak-coupling regime. Fig. 3 shows
the resulting changes. At finite $J$ the eigenfunctions in the pp{\bf 0} and
phc{\bf 0} channels are essentially nonzero all around the FS, however they
are flatter than the $d_{x^2-y^2}$-wave function. The eigenfunction in the
phs{\bf Q} channel is of extended s-wave form. The eigenvalue for the
phc{\bf 0} channel is smaller than at $J=0$, implying that stronger
antiferromagnetism weakens the tendency towards a Pomeranchuk instability.
Similarly to the $J=0$ case, we observe a smaller variation of the vertices
${\cal R}_{{\bf k}}^{m,r}$ around the FS, than in the momentum dependence of
the corresponding eigenfunctions of the Bethe-Salpeter equation.

With increasing $t^{\prime }$ to $0.3t$ (Fig. 4) the largest eigenvalue
occurs in the singlet dSC channel. The pp{\bf 0} pair wavefunction maintains
its shape with a slight deviation from the ($\cos k_x - \cos k_y$)-form,
i.e. a flattening near the BZ diagonal. In the phs{\bf Q} and phc{\bf 0}
channels we observe again a weaker momentum variation of ${\cal R}_{{\bf
k}}^{m,r}$. The vertex ${\cal R}_{{\bf k}}^{pp{\bf 0},d\text{-wave}}$ instead
is very close to the shape of the corresponding Bethe-Salpeter eigenfunctions
in the pp{\bf 0} channel.

\begin{figure}[t!]
\psfig{file=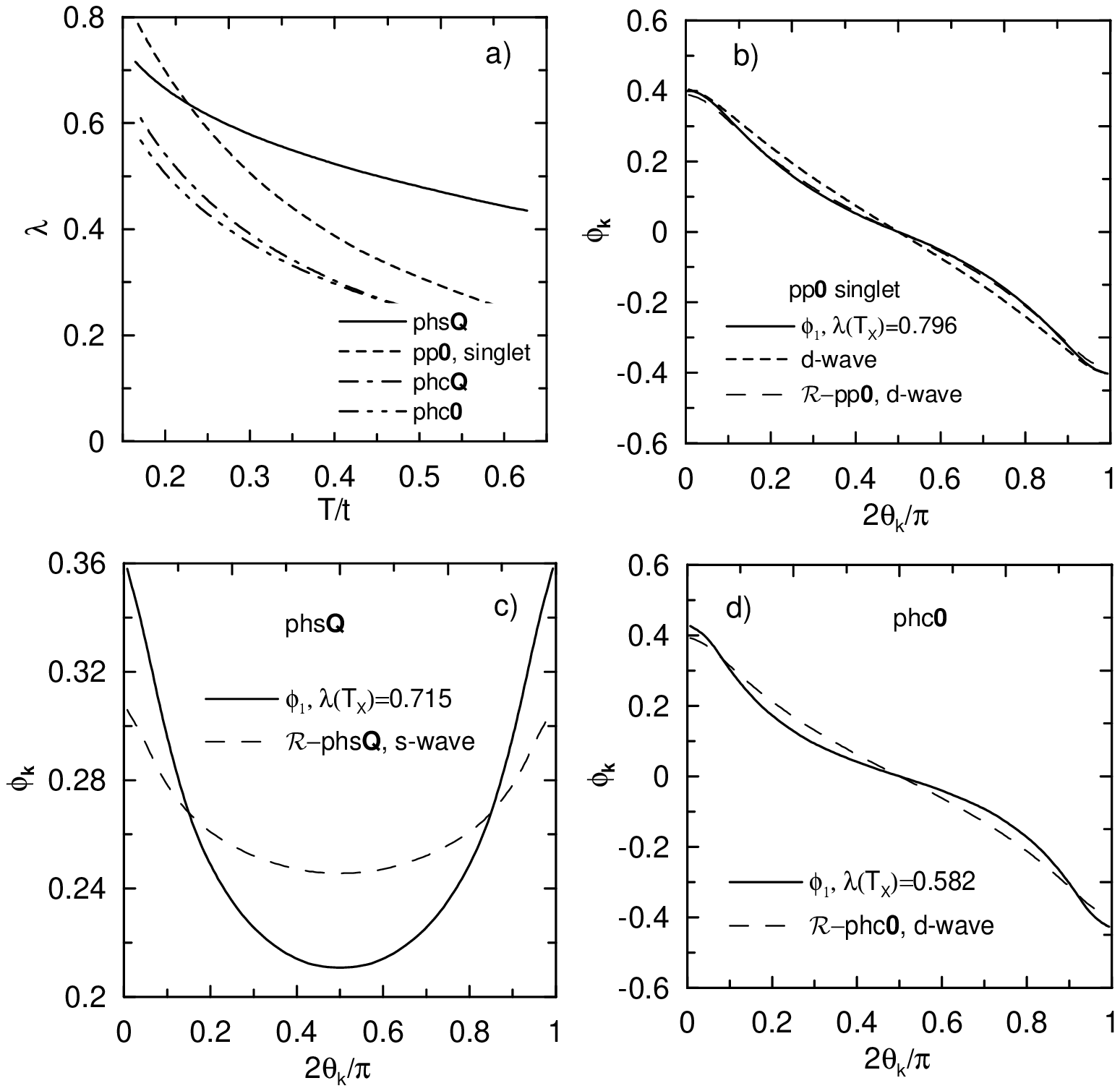,width=90mm,silent=} \vspace{2mm}
\caption{Same as in Fig. 1 for $t^{\prime }=0.3t,$ $U=2t$, $J=0.3t$. }
\label{fig:Fig4}
\end{figure}

For $t^{\prime }=0.45t$ a ferromagnetic instability is expected
\cite{SalmHon1,KK}, and we start again with the vH band filling case for
$\mu =0$. We increase the interaction strength to $U=3t$, since the
corresponding crossover temperatures for the FM instability are lower. The
largest Bethe-Salpeter eigenvalues arise in the FM (phs{\bf 0}), pSC (pp{\bf 0}
triplet) and AFM channels (phs{\bf Q}). Remarkably, the wavefunction in the
{\it triplet} pp{\bf 0} channel deviates significantly from the $\sin k_x$
form (see Fig. 5b), but its eigenvalue remains smaller than the eigenvalue
for ferromagnetism. This deviation is also well reproduced by the
corresponding momentum dependence of the vertex ${\cal R}_{{\bf k}}^{pp{\bf
0},p\text{-wave}}$.

On moving away from the vH band filling at $t^{\prime }=0.45t$, we observe a
further increase of the eigenvalue in the triplet pairing channel (Fig. 6),
which however remains smaller than the eigenvalue of the ferromagnetic
instability at the lowest temperature we can safely reach in the fRG flow.
Simultaneously,
the eigenfunction in the triplet superconducting channel slightly distorts
towards the wave function
$f_{{\bf k}}^{(p\text{-wave})}$, but strong deviations persist in both, the
eigenfunction of the Bethe-Salpeter equation and the vertex
${\cal R}_{{\bf k}}^{pp{\bf 0},p\text{-wave}}$. Singlet superconductivity,
which was not among the dominant instabilities at $\mu =0$, is also
significantly enhanced for $\mu >0$. There are two eigenfunctions, which are
symmetric and antisymmetric with respect to reflection at the BZ diagonal with
almost equal eigenvalues in the pp singlet channel. Both eigenfunctions are
essentially nonzero at ($\pi,0$) and (0,$\pi$) only, and therefore not
expected to be stabilized thermodynamically. Because of the presence of two
nearly-degenerate eigenfunctions, the three-point vertex ${\cal R}_{{\bf
k}}^{pp{\bf 0},d\text{-wave}}$ in this case substantially deviates from the
momentum dependence of the leading eigenfunction.

\begin{figure}[t!]
\psfig{file=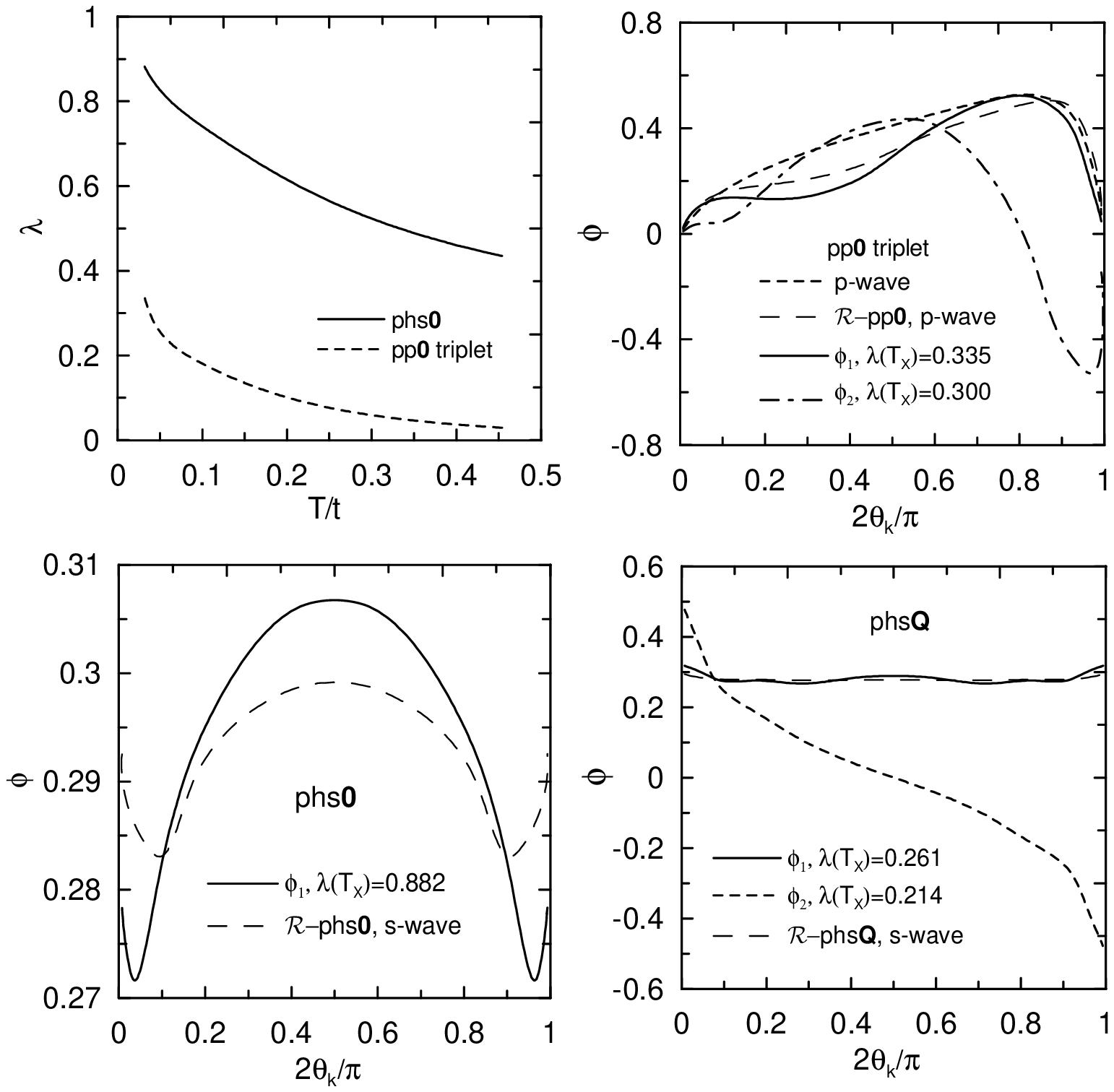,width=90mm,silent=} \vspace{2mm}
\caption{Same as in Fig. 1 for $t^{\prime }=0.45t,$ $U=3t$}
\label{fig:Fig5}
\end{figure}

In conclusion, we have investigated the symmetry of the leading
instabilities of the 2D $t$-$t^{\prime }$ Hubbard model using as a novel
tool the combination of the Bethe-Salpeter equation and the fRG approach.
Although in most cases the leading instabilities coincide with those from
susceptibility based analyses, the true shape of the eigenfunctions
significantly differs from the s-, p- or d-wave basis functions. At small
$t^{\prime }$ we find in particular a flattening of the eigenfunction in the
pairing pp{\bf 0} channel near the nodes -- in qualitative agreement with
experimental data for cuprates \cite{Mesot}. The addition of a direct spin
exchange interaction to the Hubbard model at weak-coupling was essential to
reproduce a ($\cos k_x-\cos k_y$)-like form of the superconducting gap at low
and intermediate $t^{\prime}$. The instability towards triplet pairing at
larger $t^{\prime }$ also shows a substantial deviation from the standard
p-wave $\sin k_x$-form. In many
cases we have investigated, the eigenfunctions of the Bethe-Salpeter equations
strongly deviate from the corresponding three-point vertices ${\cal R}_{{\bf
k}}^{m,r}$, which enter the order parameter susceptibilities, due to either a
qualitative difference of the eigenfunctions from the basis functions
$f_{{\bf k}}^{(s,p,d)\text{-wave}}$ or the presence of several almost
degenerate eigenfunctions. The proposed new technique may prove most useful in
future studies of magnetic and superconducting instabilities without
presupposing a special momentum dependence of the candidate order parameters.

This work was supported by the Deutsche Forschungsgemeinschaft through SFB
484. We gratefully acknowledge discussions with D. J. Scalapino, W. Metzner,
C. Honerkamp, and A. Castro-Neto.

\begin{figure}[t!]
\psfig{file=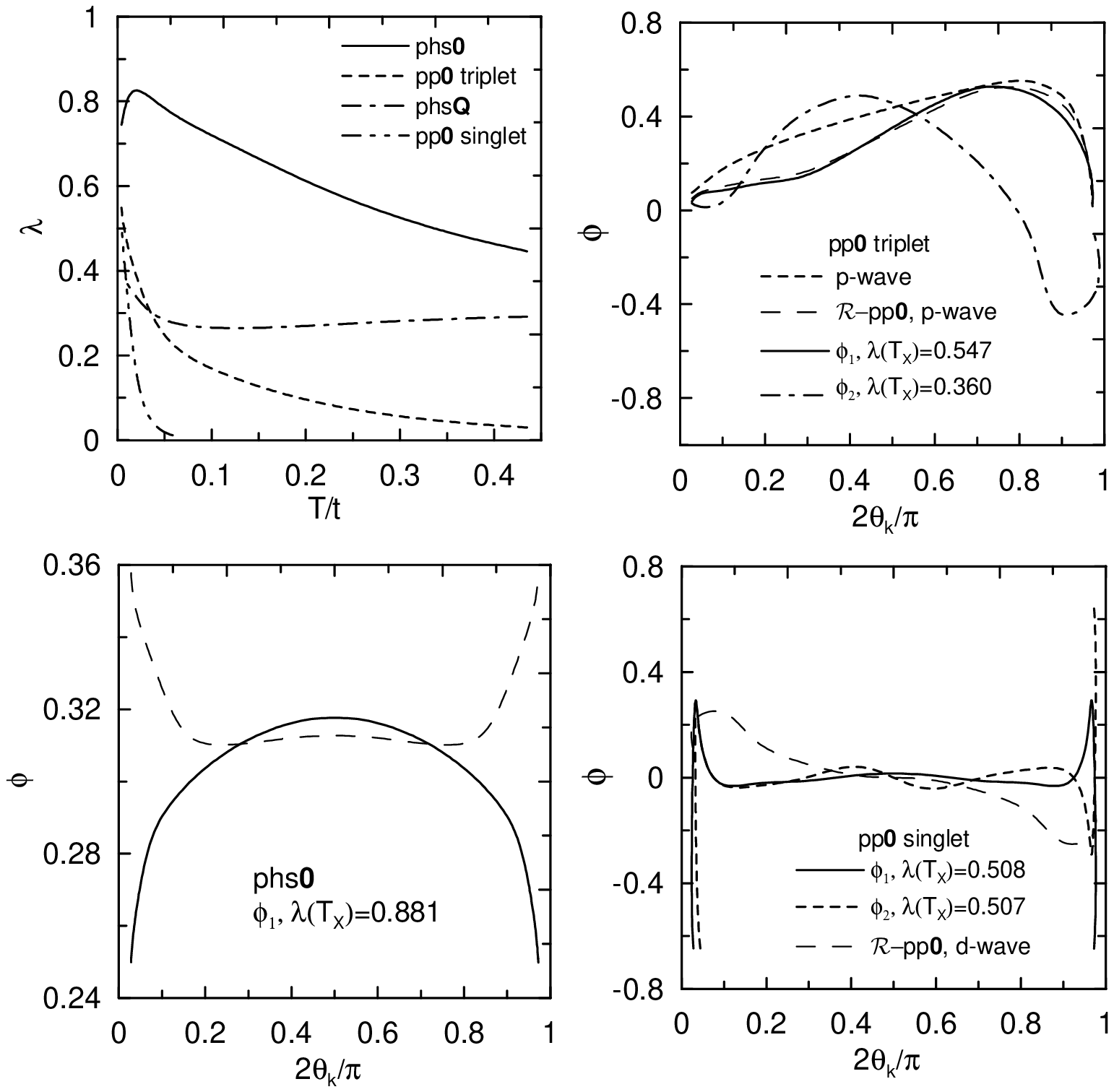,width=90mm,silent=} \vspace{2mm}
\caption{Same as in Fig. 1 for $t^{\prime }=0.45t,$ $U=3t$, and $\mu =0.035t$.}
\label{fig:Fig6}
\end{figure}

\end{document}